# Unraveling Thermally Induced Spin reorientation of Strongly Disordered NdFe$_{0.5}$Cr$_{0.5}$O$_3$ System

Jiyu Shen[1,2], Jiajun Mo[1,2], Zeyi Lu[1], Chenying Gong[1], Kaiyang Gao[1], Ke Shi[1], Lizhou, Yu[1], Yan Chen[1], Min Liu*[,1], Yanfang Xia*[,1]

[1]College of Nuclear Science and Technology, University of South China, Hengyang 421200, Hunan, P.R China.
[2]These authors contributed equally to this work.
Email: liuhart@126.com, xiayfusc@126.com

**Abstract**

Sophisticated spin instruments require high-precision spin control. In this study, we accurately study the intrinsic magnetic properties of the strongly disordered system NdFe$_{0.5}$Cr$_{0.5}$O$_3$ through molecular field models combined with ASD theory. The three constituent sub-magnetic phases of the system are separated, and their magnetization contributions are calculated separately. Fitting the angle of the A/B magnetic moment at a given temperature, the reorientation temperature point and temperature dependence of different magnetic phases are obtained. This research will provide a very good theoretical support for studying complex disordered systems and applying high-precision spin control and lay a foundation for the design of new functional materials.

**Keywords:** NdFe$_{0.5}$Cr$_{0.5}$O$_3$; spin reorientations; mössbauer; magnetic disorders; molecular field theories; anti-site defects.

**Introduction**

Rare earth perovskites are a subgroup of the larger family of perovskites. The $RFeO_3$(R: rare earth ion) type is a G-type antiferromagnetic semiconductor with a Curie temperature of about 650 K to 750 K. [1] Due to their wide range of properties and functionalities, such as ferromagnetic, giant magneto resistive, and superconducting properties, [2-4] these materials have been widely used in magneto-optical materials, [5] catalysts, [6-8] and solid oxide fuel cell sensors. [9,10]

Compared with general perovskites, the A-sites of rare-earth perovskites are occupied by rare-earth ions with unique magnetic properties. Therefore, we consider both B-site and A-site ions in the magnetic exchange model in this system. Many researchers usually use doping to change its properties, such as structure, magnetism, etc. This makes the whole system have six exchange effects of AA, $AB_1$, $AB_2$, $B_1B_2$, $B_1B_1$, and $B_2B_2$. Surprisingly, when the B-site ion ratios are very close, anti-site defects tend to appear, which is described by the anti-site defect factor (ASD). [11] At this time, the long-range magnetic order of the whole system is broken, and the entire system falls into a state of intense disorder. At this time, analysing the macroscopic magnetism of the system will be particularly complicated. Before this, my colleagues conducted a detailed theoretical study of this. [12]

Spin ultra-high-speed control has always been one of the leading research directions of spintronic and nanoscale spin-switching principles. [13] It is the basis for ultra-precise control in the future. [14] According to previous research, the control methods commonly used at present are primarily variable temperature control and femtosecond laser control. [15-17] However, unlike femtosecond laser control, the control of reorientation by means of variable temperature is generally very rough, the process of spin reorientation can only be understood qualitatively. [18,19] Rare earth perovskites are one of the materials that have the potential to achieve temperature-controlled spin. [20-22] Especially for the possibility of reorientation in a large temperature range, the research and analysis of its accurate control are particularly important. [20]

$NdFe_{0.5}Cr_{0.5}O_3$ belongs to a strongly disordered system with a reorientation temperature in the range of 70 K to 150 K. [20] In this report, we will use the molecular field model combined with Mössbauer spectroscopy to determine the magnetic disorder of $NdFe_{0.5}Cr_{0.5}O_3$ The system performs magnetic phase separation and a thorough and detailed study of the reorientation process of each magnetic phase. This study will make an instructive contribution to the magnetization analysis of strongly disordered rare-earth perovskites and provide a theoretical basis for artificial spin control of rare-earth perovskites in the later stage.

**Experimental**

Sample Preparation

$NdFe_{0.5}Cr_{0.5}O_3$ is prepared by a simple sol-gel combustion method. The required precursors are all from Macklin company, namely neodymium oxide ($Nd_2O_3$), ferric nitrate nonahydrate ($Fe(NO_3)_3·9H_2O$), chromium nitrate nonahydrate ($Cr(NO_3)_3·9H_2O$), nitric acid ($HNO_3$), ethylene glycol ($C_2H_6O_2$), citric acid ($C_6H_8O_7$).

First, the neodymium oxide was dissolved with excess nitric acid to obtain a neodymium nitrate solution. Then, other nitric acid compounds in water were mixed with neodymium nitrate solution according to a certain ionic ratio ($Nd^{3+}$: $Fe^{3+}$: $Cr^{3+}$=1: 0.5: 0.5). An excess of citric acid and a certain amount of ethylene glycol were combined as the mixed solution molar ratio of 1:1. The solution was stirred evenly at 80°C on a magnetic stirrer until a gel formation. Then, the gel was heated at 120°C. The obtained powder was pre-calcined at 600 °C for 12 hours and subsequently calcined at 1200 °C for 24 hours, upon cooling, the final nano powder samples were obtained.

X-ray diffraction measurements (XRD)

X-ray diffraction (XRD) experiments were carried out in Siemens D500 Cu Kα ($\lambda$ = 1.5418 Å) diffractometer with the range of 20° to 80° (rate of 0.02°/s). The obtained data were processed with Fullprof software.

Fourier infrared spectroscopy measurements (FT-IR)

The Fourier infrared spectrometer model L1600400 Spectrum Tow PerkinElmer were used. We performed an infrared test on the $NdFe_{0.5}Cr_{0.5}O_3$ with a wavenumber in the range of 400 $cm^{-1}$ to 4000 $cm^{-1}$.

Mössbauer spectrum Test

The transmission $^{57}Fe$ Mössbauer spectra of $NdFe_{0.5}Cr_{0.5}O_3$ were collected at RT on SEE Co W304 Mössbauer spectrometer with a $^{57}Co/Rh$ source in transmission geometry equipped in a cryostat (Advanced Research Systems, Inc.,4 K). The data results were fitted with MössWinn 4.0 software.

Magnetic Test

The thermomagnetic curves (M-T) of the samples NdFe$_{0.5}$Cr$_{0.5}$O$_3$ were obtained under an external magnetic field of 100 Oe. Field-cooled (FC) and zero-field-cooled (ZFC) measurements were done at a magnetic field strength of 100 Oe and temperatures ranging from 5 to 400 K. The magnetization curves (M-H) are obtained at temperatures of 300 K and 5 K, respectively.

**Results and discussions**

XRD

analysis

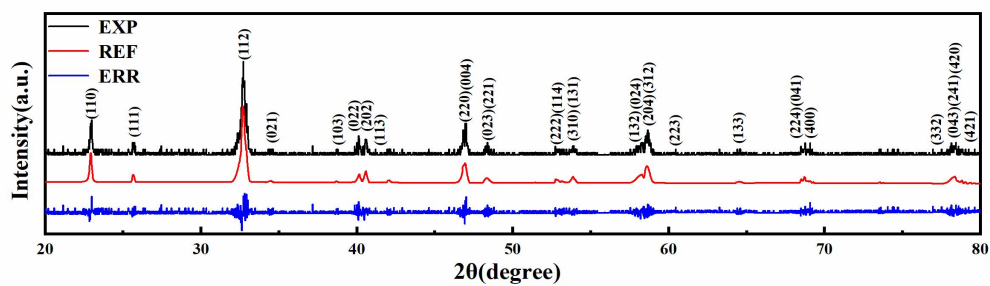

**Fig. 1** NdFe$_{0.5}$Cr$_{0.5}$O$_3$ XRD spectrum refined by Fullprof software (the black/red/blue lines represent experimental values/refinement values/errors, respectively).

**Table. 1** Lattice parameters, unit cell volume and other parameters of NdFe$_{0.5}$Cr$_{0.5}$O$_3$.

| Sample | a (Å) | b (Å) | c (Å) | Cell volume (Å$^3$) | Average grain size (Å) | Bulk density (g/cm$^3$) |
|---|---|---|---|---|---|---|
| NdFe$_{0.5}$Cr$_{0.5}$O$_3$ | 5.425 | 5.478 | 7.694 | 229.65 | 537 | 7.0946 |

The XRD results of NdFe$_{0.5}$Cr$_{0.5}$O$_3$ sample were analysed using FullProf software, and the results are presented in Figure. 1. The lattice parameters, unit cell volume, average grain size, and other parameters of the NdFe$_{0.5}$Cr$_{0.5}$O$_3$ samples can be seen in Table. 1. The prominent diffraction peaks of the samples are marked on their respective signals. It can be judged from the main Bragg peaks that all samples have orthogonal space groups. Besides, the XRD pattern shows a slight peak splitting phenomenon at 32°-33.5°. This shows that the perovskite structure undergoes orthogonal distortion under the stress contribution. The structural parameter ($D_{OD}$) is introduced to characterize the orthogonal distortion in the following formula: [23]

$$D_{OD} = \frac{1}{3} \sum_{i=1}^{3} |\frac{\alpha_i - \bar{\alpha}}{\bar{\alpha}}| * 100\% \quad (1)$$

For the lattice of the *Pbnm* space group, $\alpha_1 = a$, $\alpha_2 = b$, $\alpha_3 = c$, $\bar{\alpha} = (a * b * c / \sqrt{2})^{\frac{1}{3}}$. [23]

The Rietveld refinement using FullProf software was implemented to determine suitable lattice parameters, unit cell volume and other structural parameters. In addition, it roughly estimates the degree of dislocation of each ion in the lattice known as anti-site defect (ASD). [24] In NdFe$_{0.5}$Cr$_{0.5}$O$_3$, we estimate that the ASD of Fe$^{3+}$ ions to Cr$^{3+}$ ions are around 45%, which will be useful for us to make accurate analysis in combination with the previous theories. [12]

FT-IR analysis

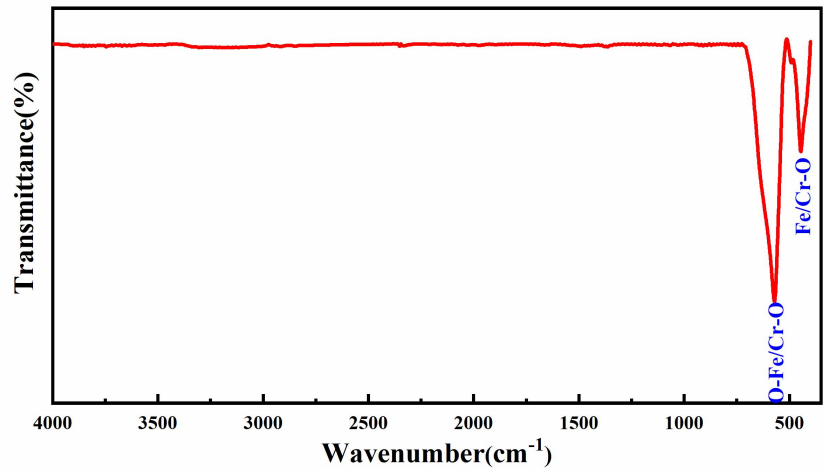

**Fig. 2** FT-IR spectrum of NdFe$_{0.5}$Cr$_{0.5}$O$_3$.

The FT-IR spectra of NdFe$_{0.5}$Cr$_{0.5}$O$_3$ is shown in Figure. 2. The sample show two distinct characteristic absorption peaks, located between 400 cm$^{-1}$-500 cm$^{-1}$ and 500 cm$^{-1}$-600 cm$^{-1}$, respectively. The absorption peaks in the lower wavenumber range are caused by the vibration of the O-Fe/Cr-O bond. The absorption peaks in the higher wavenumber range correspond to the vibration of the Fe/Cr-O bond. [25]

Mössbauer spectra

analysis

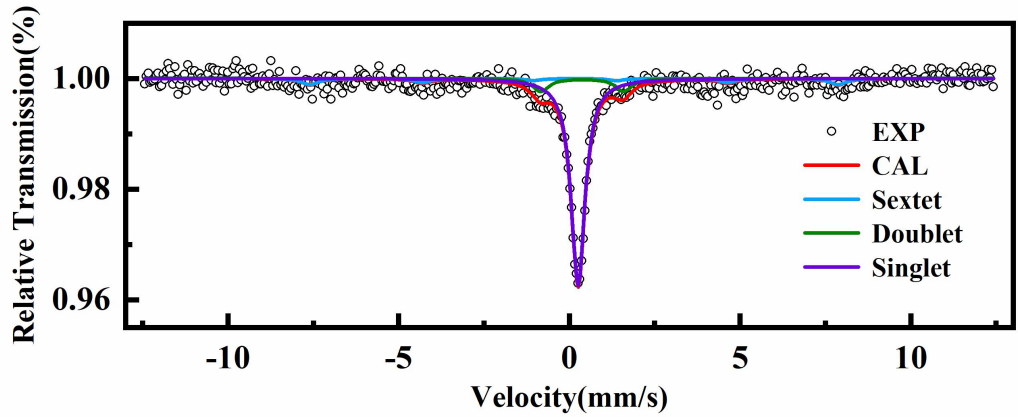

**Fig. 3** Mössbauer spectrum of NdFe$_{0.5}$Cr$_{0.5}$O$_3$ at 300 K.

**Table. 2** Mössbauer hyperfine parameters of NdFe$_{0.5}$Cr$_{0.5}$O$_3$.

| Sample | T (K) | Phase | QS (mm/s) | IS (mm/s) | H (T) | Γ (mm/s) | Area (%) |
|---|---|---|---|---|---|---|---|
| NdFe$_{0.5}$Cr$_{0.5}$O$_3$ system | 300K | Sextet | -0.032 | 0.157 | 48.081 | 0.582 | 11.1 |
| | | Doublet | 2.429 | 0.380 | / | 0.582 | 11.6 |
| | | Singlet | / | 0.257 | / | 0.491 | 77.3 |

The Mössbauer spectrum of NdFe$_{0.5}$Cr$_{0.5}$O$_3$ was measured at 300 K and parameters such as isomer shift (IS), quadrupole splitting (QS) and hyperfine field (H) were fitted using MössWinn 4.0 software. Table. 2 contains all fitted data. Figure. 3 shows the Mössbauer spectral fitting results. IS and QS values indicated the presence of Fe$^{3+}$, while no divalent or tetravalent iron was found. Three lines appear, indicating that three or more phases are expected in this compound. It is reasonable to speculate that the singlet with the largest proportion represents the NdFe$_{0.5}$Cr$_{0.5}$O$_3$ phase. And that sextet may be NdFe$_{1-p}$Cr$_p$O$_3$ ($p < 0.5$), whose phase transition temperature is higher than 300 K. Another doublet may be NdFe$_{1-q}$Cr$_q$O$_3$ ($q>0.5$), whose phase transition temperature is lower than 300 K. According to the absorption area of the two, it can be inferred that the composition ratio is the same, that is to say, $p = 1-q$.

Magnetic analysis

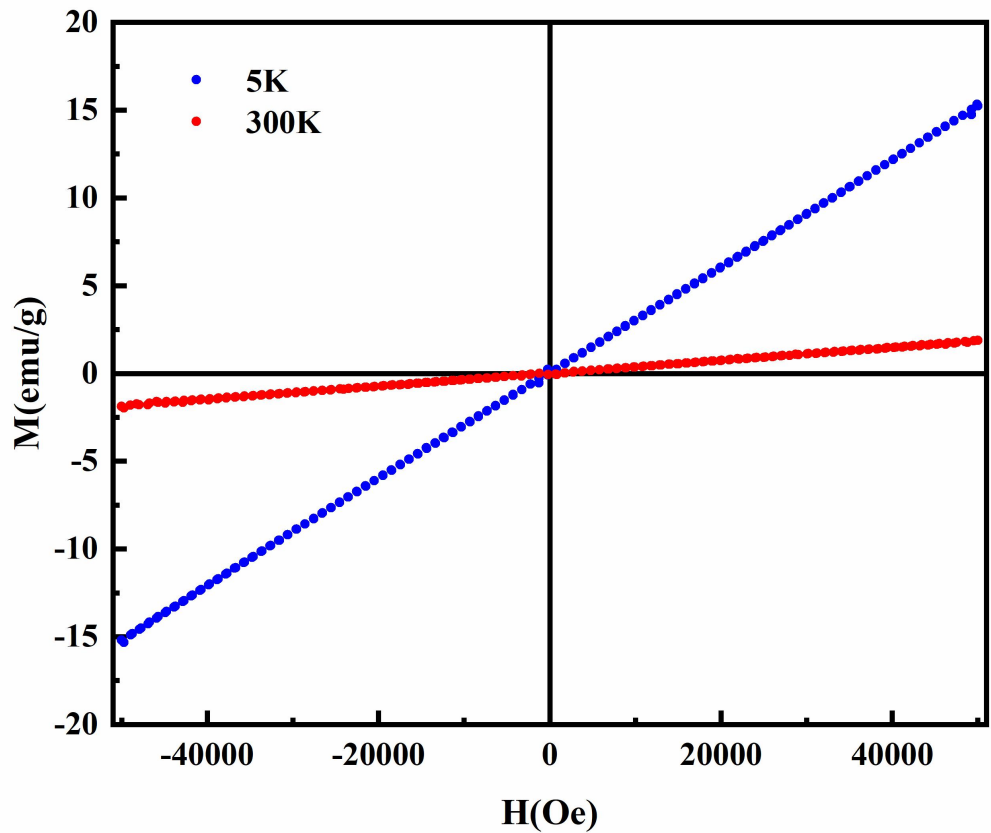

**Fig. 4** M-H curves of NdFe$_{0.5}$Cr$_{0.5}$O$_3$ tested at 5K and 300K, respectively.

We tested the magnetization curve of NdFe$_{0.5}$Cr$_{0.5}$O$_3$ system at 5K and 300K temperature. Both appear as straight lines of different slopes and do not reach saturation magnetization. It is shown that the system maintains antiferromagnetic properties at both 5 K and 300 K.

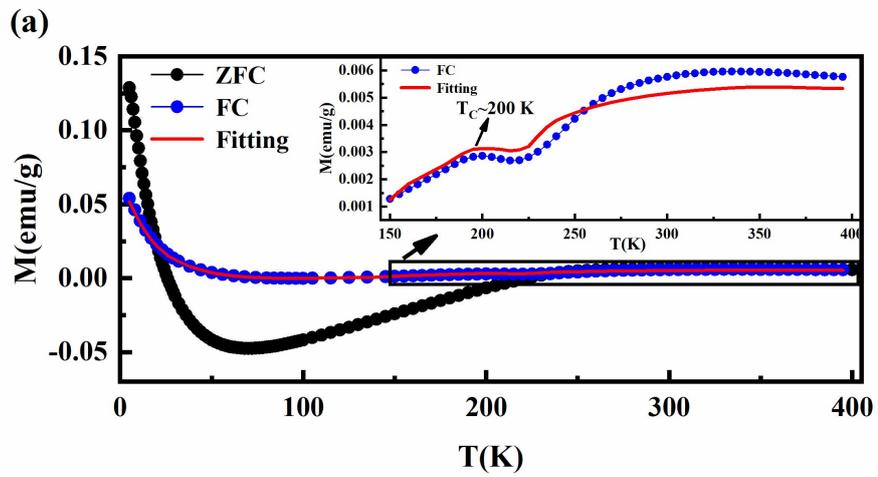

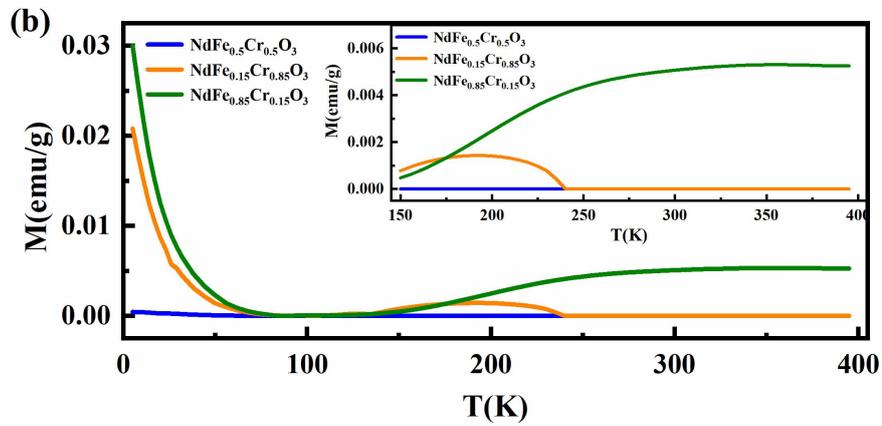

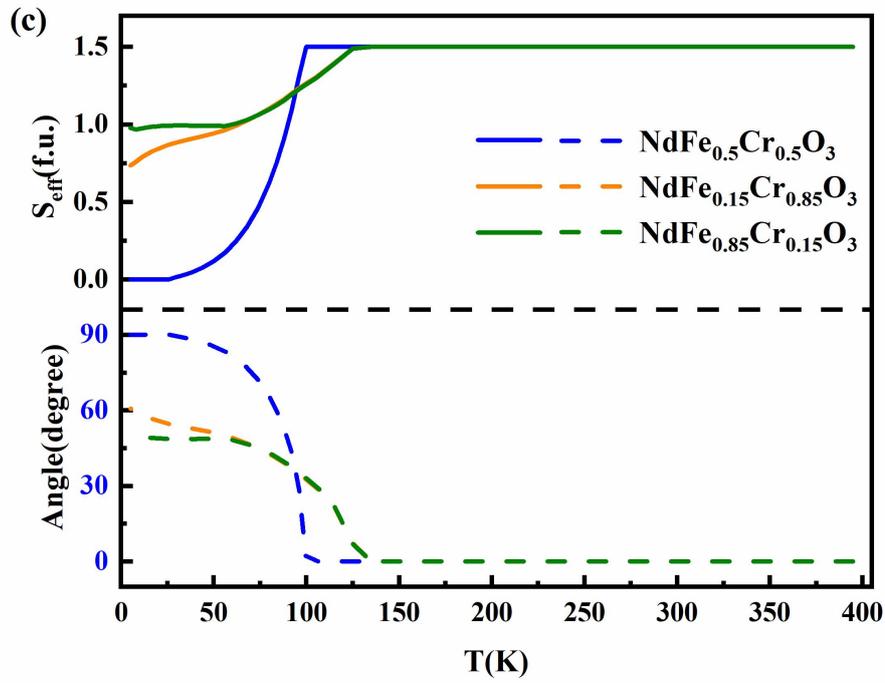

**Fig. 5** Magnetic test and fitting results of NdFe$_{0.5}$Cr$_{0.5}$O$_3$, (a) M-T curves and fitted FC curves under an external field of 100 Oe (the inset is an enlarged view of the 150 K-400 K range); (b) FC magnetization contribution curves of the three magnetic phases (inset is the 150 K- Enlarged image in the range of 400 K); (c) spin reorientation process (the upper image is the change of the effective magnetic moment, and the lower image is the change of the angle of the magnetic moment).

The macroscopic magnetism of NdFe$_{0.5}$Cr$_{0.5}$O$_3$ was measured under an external field of 100 Oe, and the results are shown in Figure. 5 (a). An obvious phase transition temperature (T$_C$) around 200K can be found through the enlarged image. This is due to the superexchange interaction caused by the remaining electron and spin-orbit coupling at this temperature point, the spin angle that promotes the reduction of Cr$^{3+}$ ions are slightly less than 180°, and weak ferromagnetic behavior is recorded below this temperature. [26] Above this temperature point, we find that the magnetization continues to rise, corresponding to the magnetization behavior of another multiferromagnetic phase, and the Curie temperature point of this magnetic phase is much larger than 200 K. In order to obtain the intrinsic magnetization characteristics of the sample, we combined the previous work--ASD theory with the molecular field model, and obtained the thermomagnetic curve reflecting its essence through three-phase fitting, as shown in the red fitting line in Figure. 5 (a).

Molecular field model and ASD theory Fitting

Based on our previous work, in the $RFe_{0.5}Cr_{0.5}O_3$ system, the order of Fe and Cr arrangement can be described by two parameters ($x$ and $y$) whose values depend on the saturation magnetization intensity Ms and the phase transition temperature $T_P$ []. The anti-site degree (ASD) of the system corresponds to $x/(x+y)$. where $x$ or $y$ is equal to zero. When $x = 0$ and $y = 0$, Anti-site atoms do not exist essentially. So $Fe^{3+}$-O-$Fe^{3+}$ and $Cr^{3+}$-O-$Cr^{3+}$ exchanges can be ignored. When $x \to 3$ and $y \to 3$, the arrangement tends to be disordered, corresponding to the case of uniform distribution. When $x$ and $y$ are equal and very small, the arrangement of atoms shows an orderly trend. In spite of this, its anti-site defect can still reach 0.5. When $x \to 6$ and $y \to 6$, the Fe-rich and Cr-rich phenomena can be evident. We can even say that the system contains two independent phases with $Fe^{3+}$ and $Cr^{3+}$ occupying one side. We mark the Fe ion lie in $a$-site as $Fe^a$, the same is true for $Fe^b$, $Cr^a$, $Cr^b$. Additionally, $Z_{ij}$ is the number of $j$ neighbor $i$. $Z_{Fe^a Fe^b} = Z_{Cr^b Cr^a} = x$, $Z_{Fe^b Fe^a} = Z_{Cr^a Cr^b} = y$, $Z_{Fe^a Cr^b} = Z_{Cr^b Fe^a} = 6-x$, $Z_{Fe^b Cr^a} = Z_{Cr^a Fe^b} = 6-y$. The mean-field can be used to express the field acting on each sublattice

$$H_{Fe^a} = \lambda_{Fe^a Fe^b} M_{Fe^b} + \lambda_{Fe^a Cr^b} M_{Cr^b} + \lambda_{Fe^a Nd} M_{Nd} + h$$
$$H_{Fe^b} = \lambda_{Fe^b Fe^a} M_{Fe^a} + \lambda_{Fe^b Cr^a} M_{Cr^a} + \lambda_{Fe^b Nd} M_{Nd} + h$$
$$H_{Cr^b} = \lambda_{Cr^b Fe^a} M_{Fe^a} + \lambda_{Cr^b Cr^a} M_{Cr^a} + \lambda_{Cr^b Nd} M_{Nd} + h \quad (2)$$
$$H_{Cr^a} = \lambda_{Cr^a Cr^b} M_{Cr^b} + \lambda_{Cr^a Fe^b} M_{Fe^b} + \lambda_{Cr^a Nd} M_{Nd} + h$$
$$H_{Nd} = \lambda_{NdFe} M_{Fe} + \lambda_{NdCr^b} M_{Cr^b} + \lambda_{NdCr^a} M_{Cr^a} + h$$

Where $\lambda_{ij}$ represents the molecular field constant between $i$ and $j$ sublattices, and it is proportional to the exchange constant $J_{ij}$, $h$ is the external field; $M_i$ is the magnetization of $i$ sublattice. The magnetization of $i$ sublattice is:

$$M_i = x_i N_A g \mu_B S_i B_{Si}\left(\frac{g \mu_B S_i H_i}{k_B T}\right) \quad (3)$$

Where $x_i$ is the molar quantity of $i$ ions, B is the Brillouin function

$$B_J(\gamma) = \frac{2J+1}{2J}\coth\left(\frac{2J+1}{2J}\gamma\right) - \frac{1}{2J}\coth\left(\frac{\gamma}{2J}\right),$$ $g$ is the lande factor, and $\mu_B$

represents the Bohr magneton. $N_A$ is the Avogadro constant. $S_i$ is the spin quantum number of $i$ ions ($S_{Fe}$ = 5/2, $S_{Cr}$ = 3/2, $S_{Nd}$ = 3/2). The exchange constant $J_{MM}$ between $M$ and $M$ ions can be calculated by

$$|J_{MM}| = \frac{2Z_{MM}S_M(S_M+1)}{3k_B T_N^M} \tag{4}$$

Where in $Z_{MM}$ is the number of $M$ ions required to be $M$ ions nearest neighbors. $k_B$ is the Boltzmann constant, and $T_N^M$ is the Phase transition temperature of NdMO$_3$. The exchange constant that $J_{FeFe}/k_B$ is about -22 K and $J_{CrCr}/k_B$ is about -16.6 K for $T_N$ of NdFeO$_3$ ~ 660 K and that of NdCrO$_3$ ~ 250 K.

Previously, we obtained through Mössbauer spectroscopy that the complex system of NdFe$_{0.5}$Cr$_{0.5}$O$_3$ not only has a phase of NdFe$_{0.5}$Cr$_{0.5}$O$_3$, but also has a ferromagnetic state and another phase in a paramagnetic state at 300 K. By changing the Fe/Cr ratio of the two ordered phases to match the Curie temperature point of 200 K in the experimental data, we obtained a ratio of approximately NdFe$_{0.15}$Cr$_{0.85}$O$_3$ and NdFe$_{0.85}$Cr$_{0.15}$O$_3$, that is, NdFe$_{0.15}$Cr$_{0.85}$O$_3$ The corresponding local temperature point is 200 K. Thus, we replace the complex system with three magnetic phases: NdFe$_{0.5}$Cr$_{0.5}$O$_3$, NdFe$_{0.15}$Cr$_{0.85}$O$_3$, NdFe$_{0.85}$Cr$_{0.15}$O$_3$.

By XRD we have obtained an ASD value of about 45% for NdFe$_{0.5}$Cr$_{0.5}$O$_3$, which translates to $x$ = 3.5413, $y$ = 4.3353. The other two magnetically ordered phases have an ASD of 0%, which translates to $x$ = 5.99999, $y$ = 4.80001. We first tried to directly fit the experimental curve with the ASD of NdFe$_{0.5}$Cr$_{0.5}$O$_3$, and the best result found that it has no magnetic contribution at high temperature. From this, we can separate the magnetic contribution of NdFe$_{0.85}$Cr$_{0.15}$O$_3$ separately, and simplify the problem to the two-phase fitting of NdFe$_{0.5}$Cr$_{0.5}$O$_3$ and NdFe$_{0.15}$Cr$_{0.85}$O$_3$, which solves the problem caused by parameterization.

The final result is shown in Figure 5 (b). It can be seen that although the content of NdFe$_{0.85}$Cr$_{0.15}$O$_3$ is very small, the magnetization contribution accounts for the largest proportion. In contrast, the magnetization contribution of NdFe$_{0.5}$Cr$_{0.5}$O$_3$ only has a small part in the ultra-low temperature section. The phase transition of NdFe$_{0.15}$Cr$_{0.85}$O$_3$ occurs unexpectedly at 200 K and ends at 235 K. The respective exchange constants of the three magnetic phases are shown in Table. 3.

In the fitting process, we consider the magnetization of NdFe$_{0.5}$Cr$_{0.5}$O$_3$ as a vector superposition of Fe$^{3+}$/Cr$^{3+}$ and Nd$^{3+}$ ions based on the spin reorientation process considered. Since the A/B sites have different easy axis orientations, their interactions can be determined by their spin projections on specific planes. We believe that the interaction plane can be ascribed to the Fe/Cr spins, so we were able to demonstrate how S$_{Nd}$ changes with temperature. S$_N$ is defined as the projection of the Nd$^{3+}$ ion spin on the plane containing the Fe/Cr spin. The exchange constants are fixed across the temperature range, solving for the S$_N$ at each temperature. The results are shown in Figure. 5 (c). It can be seen that the spin reorientation process of each magnetic phase is that the effective magnetic moment decreases with the decrease of temperature, but the effective magnetic moment at the initial temperature and the lowest temperature are different. $NdFe_{0.5}Cr_{0.5}O_3$ undergoes spin reorientation at about 100 K, and the effective magnetic moment becomes 0 at the lowest temperature (that is, the angle between $Nd^{3+}$ and the magnetic moment of the $Fe^{3+}/Cr^{3+}$ plane turns 90°). While, the reorientation processes of $NdFe_{0.15}Cr_{0.85}O_3$ and $NdFe_{0.85}Cr_{0.15}O_3$ are very close, both occur at around 125 K, and the final effective magnetic moment is also fixed between 0.75 and 1 (that is, the angle between the magnetic moment of $Nd^{3+}$ and $Fe^{3+}/Cr^{3+}$ plane turns 50° to 60°). The overall spin reorientation process is represented in Figure. 6. We can think that the reorientation temperature may be related to the doping ratio of the B site, and the difference in the reorientation process of $NdFe_{0.15}Cr_{0.85}O_3$ and $NdFe_{0.85}Cr_{0.15}O_3$ may be due to a slight difference between the predicted ratio and the actual ratio. All algorithms in this study use the Marine Predator Algorithm (MPA). [27]

Table. 3 The exchange constants of the three magnetic phases in the $NdFe_{0.5}Cr_{0.5}O_3$ system.

| Sample | $J_{Nd-Fe}$ (K) | $J_{Nd-Cr}$ (K) | $J_{Fe-Fe}$ (K) | $J_{Cr-Cr}$ (K) | $J_{Fe-Cr}$ (K) |
|---|---|---|---|---|---|
| $NdFe_{0.5}Cr_{0.5}O_3$ | -5.558 | -5.556 | -22 | -16.571 | -8.324 |
| $NdFe_{0.15}Cr_{0.85}O_3$ | -5.971 | -5.556 | -22 | -16.571 | -8.989 |
| $NdFe_{0.85}Cr_{0.15}O_3$ | -6.025 | -5.556 | -22 | -16.571 | -9.014 |

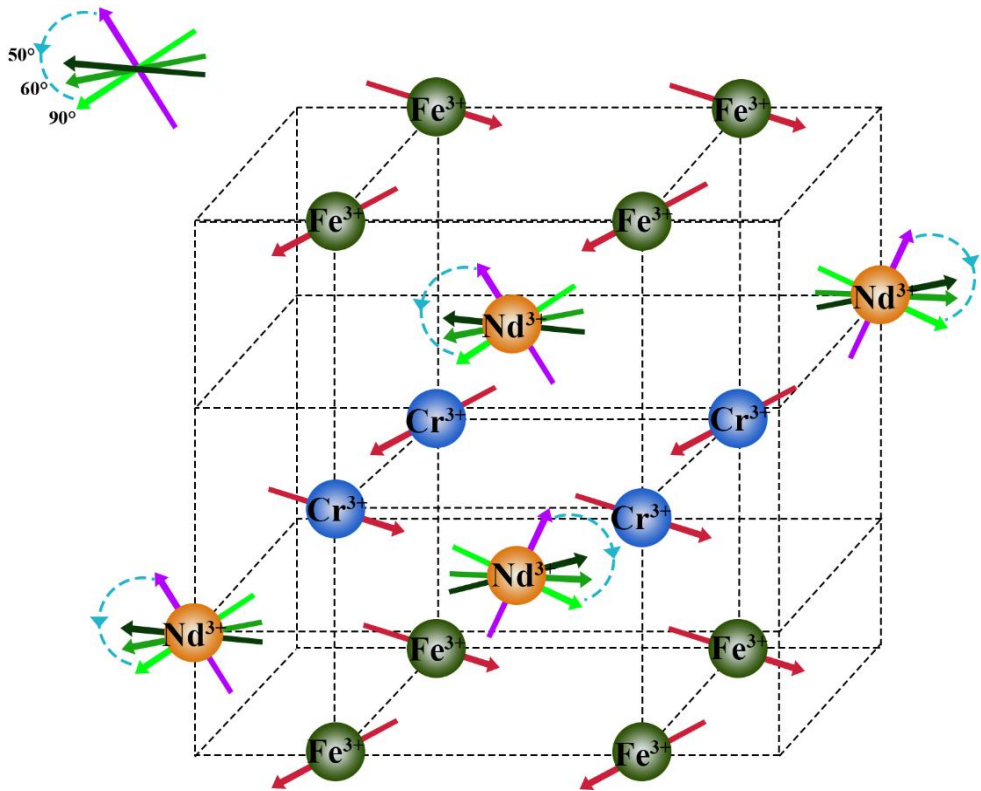

**Fig. 6** Schematic diagram of the spin reorientation process of NdFe$_{0.5}$Cr$_{0.5}$O$_3$.

**Conclusions**

In summary, we use the four-sublattice molecular field model combined with our previously proposed ASD theory to analyze the intrinsic magnetic properties of the NdFe$_{0.5}$Cr$_{0.5}$O$_3$ complex strongly disordered system. Based on the hyperfine analysis of the Mössbauer spectrum, we divide the system into three phases and fit the respective magnetization contributions, and the results are satisfactory. Not only that, in the process of fitting, the effective magnetic moment is defined as the projection of the Nd$^{3+}$ magnetic moment on the Fe$^{3+}$/Cr$^{3+}$ magnetic moment plane, and the spin reorientation process of each magnetic phase is analyzed to obtain an accurate, effective magnetic moment concerning temperature—the process of change. The temperature of reorientation is strongly related to the proportion of B-site ions. This work is different from the qualitative analysis of thermally induced spin reorientation of simple single crystal structures by previous researchers and obtained a very real dependence relationship, which has a profound guiding role in the analysis of complex disordered systems. And provide theoretical support for ultrafast spin control.

**Conflicts of interest**

There are no conflicts to declare.

**Acknowledgements**

This work was supported by National Natural Science Foundation of China (grant number 12105137, 62004143), the Central Government Guided Local Science and Technology Development Special Fund Project (2020ZYYD033), the National Undergraduate Innovation and Entrepreneurship Training Program Support Projects of China, the Natural Science Foundation of Hunan Province, China (grant number S202110555177), the Natural Science Foundation of Hunan Province, China (grant number 2020JJ4517), Research Foundation of Education Bureau of Hunan Province, China (grant number 19A433, 19C1621).